\documentstyle[12pt]{article}
\newcommand{\eq}[1]{(\ref{#1})}
\newcommand{\diff}{\partial}
\newcommand{\beq}{\begin{equation}}
\newcommand{\eeq}{\end{equation}}
\newcommand{\beqn}{\begin{eqnarray}}
\newcommand{\eeqn}{\end{eqnarray}}
\newcommand{\cD}{{\cal D}}
\def\cA{{\cal A}}
\def\cS{{\cal S}}

\def\cK{{\cal K}}
\def\cN{{\cal N}}
\def\cR{{\cal R}}
\def\cF{{\cal F}}

\def\cO{{\cal O}}

\def\NP{ Nucl.~Phys.}
\def\PR{ Phys.~Rev.}
\def\PL{ Phys.~Lett.}
\def\PRL{ Phys.~Rev.~Lett.}

\begin{document}

\hfill {\bf ITEP-TH-32/98}

\hfill hep-th/9806217
\vspace{10mm}


\centerline{\bf \Large A remark on the AdS/CFT correspondence}
\centerline{\bf \Large and the renormalization group flow.}

\vspace{5mm}

\centerline{E.T.~Akhmedov\footnote{e--mail: akhmedov@vxitep.itep.ru}}

\centerline{Institute of Theoretical and Experimental Physics}

\centerline{Moscow, 117259, B. Cheremushkinskaya, 25.}

\vspace{5mm}

\begin{abstract}
The correspondence between the four-dimensional $SU(N)$, $\cN = 4$ SYM taken at large $N$
and the type II B SUGRA on the $AdS_5\times S_5$ background is considered.
We argue that the classical equations of motion in the SUGRA picture can be
interpreted as that of the renormalization group on the SYM side. In fact,
when the D3-brane is slightly excited higher derivative terms in the field
theory on its world-volume deform it form the conformal $\cN = 4$
SYM limit.  We give arguments in favor of that the deformation goes in the
way set by the SUGRA equations of motion. Concrete example of the $s$-wave
dilaton is considered.
\end{abstract}


\vspace{5mm}

1.Recent developments in super-string theory \cite{rewdu1,rewdu2,rewdu3}
indicate that one can incorporate in it unusual non-perturbative excitations.
The latter are sub-manifolds of the ten-dimensional space-time bulk on which
strings can terminate \cite{Pol,Hor} -- so called D-branes. They appear to be
a new useful tool for studying low energy dynamics in different SUSY {\it
field} theories \cite{rewdu1}-\cite{rewdu3}.  Remarkably, the D-branes give a
geometric description of different phenomena in the SYM theories which live
on their world-volumes \cite{Wit95}-\cite{Geom3}. For example, movement in
flat directions, i.e. the Higgs mechanism, is represented as splitting and
joining of the D-branes \cite{Wit95}. Thus, masses of different fields
and couplings in the SYM and SQCD theories are represented as distances
between D-branes and angles of their respective orientation
\cite{Geom1}-\cite{Geom3}.

   Having in mind those facts, in this note we try to give a geometric
interpretation of the renormalization group flow in the SYM theory. Our work
is based on the proposed duality \cite{Mal97} between the four-dimensional
large $N$ SYM theory and type IIB SUGRA on a background which we describe
below.  Concretely, we argue that the classical equation of motion of the
dilaton in the bulk SUGRA theory is nothing but the renormalization group
equation for the SYM "coupling constant" on the D-brane world-volume
\cite{Anton}. As we discuss below, this fact is sensible if the dilaton --
"coupling constant" -- is excited and, hence, is a function of the
four-dimensional coordinates.  While if the dilaton is an arbitrary constant
it remains to be the one at any energy scale. In the both cases the movement
in the direction transversal to the D-brane is the renormalization group
transformation \cite{Mal97,DoPoSt} in the field theory on its world-volume.

   In fact, it is widely believed that super-strings suggest the
regularization of field theory \cite{GrScWi}. What is new in the
D-brane case is that the regularization of their world-volume field theories
can happen at much smaller energy scale than the Plank one, if some
particular double scaling limit is taken \cite{Anton,Kle971,Kle972}. The
regularized theory should be considered as a vacuum in that of super-strings
\cite{Mal97}.  For after the regularization we are missing information
about high frequency modes.  In our case, from the point of view of a bulk
observer they are flying away from the D-brane \cite{GuKl97}. However, in the
limit under consideration they do not escape to infinity. On the contrary,
they stay inside the throat near the black brane horizon \cite{MaSt}. So
that, from the point of view of a D-brane world-volume observer the following
is happening.  At low energy scale, smaller than the curvature of the black
brane solution, we have the SYM theory. But as we deform to bigger energy
scales higher derivative terms in the D-brane world-volume action becoming
relevant.  The latter are summing up and deform the theory from the conformal
limit. At the same time if the dilaton is excited and
becomes a function of the four-dimensional coordinates, during the deformation
in question it acquires the dependence on the energy scale.  Which, as we
argue below, is defined by the corresponding SUGRA theory.

  All that resembles the well established observations in old
matrix models. The matrix models suggest the regularization of the
two-dimensional conformal field theories with discreet target spaces.
Moreover, their "renormalization group" flows in the vicinity of the
conformal points (continuum limits) are described by equations from various
integrable hierarchies \cite{Banks,Mor}.

2.  As the beginning of our presentation, we describe here a few features of
the D-brane physics and of the correspondence between SYM and SUGRA theories.
This is done just to set the notations.

The D-branes are charged with respect to the R-R fields and preserve a
part of supersymmetry in the corresponding super-string theory \cite{Pol95}.
The {\it low} energy action describing dynamics
of one D-brane and its interaction with bulk modes looks, in the light-cone
gauge, as follows \cite{rewdu2,rewdu3,Lei89}:

\beqn
S = T_p \int d^{p+1}\xi
e^{-\varphi}\sqrt{- \det\left(g_{mn} + b_{mn} + 2\pi
F_{mn}\right)} + Q_p \int d^{p+1}V_{m_0...m_p}
C_{m_0...m_p} + \nonumber \\ + \frac{1}{2 \kappa^2}\int d^{10}
x\sqrt{-G}e^{-2\varphi}\left[\cR +
4\left(\nabla_{\mu}\varphi\right)^2 - \frac{1}{12} \left(\diff_{[\mu}
B_{\nu\alpha]}\right)^2 \right] + ... + O(\alpha'), \nonumber\\
F_{mn} = \diff_{[m}
A_{n]}, \quad g_{mn} = G_{ij}\diff_m \phi_i\diff_n \phi_j + G_{in} \diff_m
\phi_i + G_{mn}, \nonumber \\
b_{mn} = B_{ij}\diff_m\phi_i\diff_n\phi_j + B_{i[n} \diff_{m]}\phi_i +
B_{mn} \quad  i,j = p+1,...,9 \quad m,n = 0,...,p. \label{BI}
\eeqn
Where dots stand for the R-R fields and fermionic terms. In this formula
$\kappa = 8\pi^{\frac72}g_s\alpha'^2$, where $g_s$ and $\alpha'$ are the string
coupling constant and its inverse tension, respectively; $p$ is the spatial
dimensionality of the D-brane world-volume; $T_p =
\frac{\pi}{g_s} \left(4\pi^2\alpha'\right)^{\frac{3-p}{2}}$ and $Q_p$ are
its tension and charge with respect to the R-R tensor field
$C_{m_0...m_p}$.  Also $G_{\mu\nu}, B_{\mu\nu}, \varphi \quad (\mu,\nu =
0,...,9)$ are NS-NS \cite{GrScWi}
closed string modes, living in the bulk. While $\phi_i$ and $A_m$ are the
D-brane coordinates and the gauge field living on its world-volume \cite{Pol}, respectively.
The action \eq{BI} also has a generalization to the
case of $N\ge 2$ coinciding D-branes \cite{Tse}. Which describes the situation
when the open strings carry $U(N)$ Chan-Paton factors \cite{Wit95}.

    We are interested in degenerations of the theory \eq{BI}, in which the
bulk and D-brane modes seemingly decouple from each other. The first one
happens at low energies and small enough $g_s$, if the theory is considered
from the point of view of an observer, placed at a big distance from the D-brane position.
The observer does not see fluctuations of the D-brane modes ($A_m$, $\phi_i$
and their super-partners). In this case, one is left with the
ten-dimensional SUGRA containing $\delta$-functional sources of the mass and R-R
charge. The $\delta$-functions have supports on the $p+1$-dimensional
sub-manifolds. One can get reed of them via introduction of
the classical black brane background \cite{DuKh94,CaHaSt} into the SUGRA action.

 The second degeneration happens at low energies and small enough $g_s$,
if the theory is considered
from the point of view of a small distance observer. This observer does not feel long
wavelength fluctuations of the bulk modes ($G_{\mu\nu},
B_{\mu\nu}, \varphi$, R-R fields and their super-partners).
Therefore, when all other fields are
small, we get the maximally super-symmetric $U(N)$ SYM theory living on
the D-brane world-volume \cite{Wit95}.

   At first sight, the two limits in question describe
different types of theories.  It appears, though, that they coincide,
if a particular double scaling limit (with $N\to\infty$) is taken
\cite{Mal97,DoPoSt,Kle971,Kle972}.
Thus, actually the decoupling between the bulk and D-brane modes does not happen \cite{MaSt}.
Qualitatively, as an observer goes further from the D-brane position, one averages
over fluctuations in the SYM theory. The result of the averaging is the classical SUGRA on
a particular background as an effective theory for the SYM \cite{M-atrix}.

  An example of this situation is the correspondence
between the four-dimensional $\cN = 4$, $SU(N)$ SYM
theory taken at large $N$ and the type IIB SUGRA on the $AdS_5\times
\left(S_5\right)_N$ background \cite{Mal97}.
The latter contains $N$ units of the elementary flux of the R-R field
$C_{0...4}$ through the $S_5$, which is indicated by the subscript $N$.
The geometry in question \cite{Mal97}, being equal to:

\beqn
ds^2 = \frac{r^2}{R^2} \eta_{mn} dx_m dx_n +
\frac{R^2}{r^2} dr^2 + R^2 d\Omega_5, \quad n,m = 1,...,3; \nonumber \\
{\rm where} \quad r = \sqrt{x^2_4 + ...  + x^2_9}, \label{metr}
\eeqn
is valid near the D3-brane horizon $r<R$.
The horizon is at $r = 0$. Also in this correspondence one takes $g_{ym}^2 = const \cdot
g_s$ and the radii of the $AdS_5$ and $S_5$ are defined as $R^4 = 4 \pi
\alpha'^2 g_s N$. Below we are going to work with the Euclidean signature
and to represent the $AdS_5$ as: $ds^2 = \frac{R^2}{z^2}\left[dz^2 +
(d{\vec x})^2\right]$ with the boundary at $z = 0$. Where $z$ is
related to $r$ as $z = \frac{R^2}{r}$.

 The theory we get at large $N$ is the type II B non-linear $\sigma$-model
on the $AdS_5\times \left(S_5\right)_N$ background \cite{TsMe}.
As we have mentioned, it is expected \cite{Mal97} that the theory has two degeneration limits.
The first one happens when $g_{ym}^2 N \sim g_s N << 1$
and leads to the weekly coupled SYM theory. While the second degeneration happens
when $\frac{R^4}{\alpha'^2} = g_s N >> 1$ and leads to the II B SUGRA on the
$AdS_5 \times \left(S_5\right)_N$ background.
Usually it is said that the strong coupling limit
($g_{ym}^2 N \to \infty$) of the four-dimensional $SU(N)$, $\cN = 4$ SYM theory
taken at large $N$ is described by
the type II B SUGRA on the $AdS_5 \times \left(S_5\right)_N$ \cite{Mal97}.

    One might ask the following question: in what sense there is a
correspondence between the SYM and the SUGRA theories? As an answer to this
question, a more exact formulation of the statement was suggested in
\cite{GuPoKl98,Wit98}. Which establishes that as $g_s N \to \infty$ and
$N\to\infty$ we have:

\beq
<e^{-\sum_j \int J_0^j \cO^j d^4x}> \approx
e^{- I^{min}(AdS_5\times (S_5)_N)|_{J^j|_b \sim J_0^j}}. \label{boundary}
\eeq
Where on the LHS the average is taken in the strongly coupled $\cN = 4$,
$SU(N)$ SYM theory and $\left\{\cO^j\right\}$ is a complete set of
operators in it. While on the RHS $I^{min}$
is the type IIB SUGRA action minimized on classical
solutions, represented schematically as $J_j$. These
solutions have asymptotic values at the boundary $J^j|_b \sim J_0^j$
in the sense explained in \cite{Wit98}.
The latter serve as sources in the LHS.

  The belief in this correspondence is partially based on the equivalence of two
absorption probabilities of bulk modes by the D-branes \cite{Kle971,Kle972}.
One of them is computed in the large $N$ SYM picture while the other in that of the
SUGRA.

   To set the notations, we briefly describe the example of the $s$-wave dilaton
which is independent from angles of the $S_5$. From \eq{BI} we get the vertex
operator for the dilaton interaction with D-brane modes. It is equal to
$\varphi\cO_{\varphi} \sim \varphi \cdot tr \left(F_{mn}^2 + ...\right)$,
where dots stand for super-partners and higher derivative terms. Having this
vertex at our disposal, we find the absorption probability of the dilaton
by the D-brane \cite{Kle971,Kle972}.

   At the same time, in the SUGRA picture the calculation goes as follows
\cite{Kle971,Kle972}. For the $s$-wave dilaton taken as $\varphi(z,x) = (kz)^2
\chi(z) e^{i {\vec k} {\vec x}}, \quad {\rm with} \quad k = |{\vec k}|$
we get the Laplace equation in the black 3-brane background:

\beq
\left[\left(z\diff_z\right)^2 - \left(\frac{R^2}{z^2} + \frac{z^2}{R^2}\right)
(kR)^2 - 4 \right]\chi(z) = 0. \label{eqd}
\eeq
Solving the scattering problem for this equation, we
find the absorption probability when $(kR)^4 << 1$ \cite{Kle971,Kle972}.
The latter appears to be equivalent to
the absorption probability found in the SYM picture.

3.  To proceed with the main topic of the note let us examine those facts.
It is believed \cite{GuPoKl98,Wit98} that the D-brane theory is the conformal
$\cN = 4$ SYM at an energy scale smaller than $R^{-1}$. At the same time, in
the throat region ($z \ge R$) when $(kR)^2 << 1$ the equation \eq{eqd}
becomes \cite{GuPoKl98}:

\beq
\left[\left(z\diff_z\right)^2 - (kz)^2 - 4\right]\chi(z) \approx 0. \label{equ}
\eeq
Let us take the following solution of this equation:

\beq
\chi(z) = \frac12 \cK_2(kz), \quad {\rm then} \quad
\varphi (x,z) = e^{i {\vec k} {\vec x}} \frac12 (kz)^2 \cK_2(kz). \label{ren}
\eeq
Here $\cK_2$ is the modified Bessel function and as $kz\to 0$ we get
$\varphi (x,z) \approx e^{i {\vec k} {\vec x}}$. It is this
solution which is regular at the horizon ($z\to\infty$)
\cite{GuPoKl98}.

  Now, if $z \approx R$ the equation \eq{equ} becomes:

\beq
\left(z\diff_z\right)^2\chi(z) \approx 4 \chi(z), \quad {\rm then} \quad z
\diff_z \chi(z) \approx - 2 \chi(z).
\eeq
One can recognize here the renormalization group equation for the
dilaton\footnote{ Which defines the SYM coupling constant:
$\frac{1}{g_{ym}^2} \sim e^{\varphi_{\infty}}$. Here $\varphi_{\infty}$ is
the vev of the dilaton.} $\varphi \sim (kz)^2 \chi(z) \sim 1$ in the
conformal SYM theory\footnote{The equation $z\diff_z\chi'(z) = 2 \chi'(z)$
probably corresponds to the unity operator which couples to the volume
element $\cO_u \sim \sqrt{det g_{mn}}$ on the D-brane. The unity operator
behaves as $\sim z^4$, when $kz\to 0$. Which perfectly cancels the scale
dependence of the volume element as it should be in the conformal field
theory.} \cite{Anton}.  As we mentioned the latter is valid on the energy
scale smaller than $R^{-1}$.  Thus, the $z$ coordinate resembles
the normalization point in the renormalization group \cite{Mal97}.  Below we
argue that it is really the case.  Concretely, we show that a few leading
non-trivial terms in the expansion of \eq{ren} over $(kz)^2$ can be recovered
from the renormalization group in the "SYM picture".

  To begin with, we examine the relation \eq{boundary} in more details. It
is rather obscure because both sides in it are divergent. For example, the
SUGRA action on the $AdS_5$ background

\beq
I(\varphi) = \frac{\pi^3 R^8}{4 \kappa^2} \int d^4 x d z \frac{1}{z^3}\left[
\left(\diff_z\varphi\right)^2 + \left(\diff_m \varphi\right)^2
\right] \label{clac}
\eeq
has the IR divergence for the solution \eq{ren} \cite{GuPoKl98,Wit98}.

    As was argued in \cite{GuPoKl98,Wit98,WiSu98} the IR regularization on the
RHS of the \eq{boundary} is related to the UV one on the LHS.  Let us use
this observation.  There is a natural IR regularization of the \eq{clac}
\cite{GuPoKl98,Wit98}. In fact, one can shift the boundary of the $AdS_5$
from $z = 0$ to $z = \epsilon \ge R$.  The regularized action \eq{clac} in
this case is given by \cite{GuPoKl98,Wit98}:

\beqn
I_{\epsilon}^{min}(\varphi_0) \sim N^2 \int d^4 x \int d^4 y~~
\varphi_0(x) \varphi_0(y) \frac{1}{\left(\epsilon^2 + |{\vec x} - {\vec y}|^2
\right)^4} - \nonumber \\ - 2 N^2 \epsilon^2 \int d^4 x \int d^4 y~~
\varphi_0(x)\varphi_0(y) \frac{1}{\left(\epsilon^2 + |{\vec x} -
{\vec y}|^2\right)^5}, \label{regac}
\eeqn
with $\varphi_0(x) = e^{i{\vec k}{\vec x}}$.

  Now consider the generating functional in the SYM picture:

\beqn
Z(\varphi_0) = \int \cD A_{m} ... \times \exp\Bigl\{- \frac{1}{g_{ym}^2}\int
d^4 x~~tr \left(F_{mn}^2 + ...\right) + \nonumber \\ + \frac{1}{g_{ym}^2}
\int d^4 x~~\varphi_0(x) \cdot tr \left(F_{mn}^2 + ...
\right)\Bigr\}. \label{genSYM}
\eeqn
From now on dots stand for the super-partners.

  Integrating over the SYM fields in \eq{genSYM} at one loop, we get:

\beqn
Z_{\epsilon}(\varphi_0) = const \cdot \exp \Bigl\{- const\cdot
\int d^4 x \int d^4 y~~\varphi_0(x) \cdot
\varphi_0(y) \times \nonumber \\ \times <tr\left(F_{lm}^2(x)+ ...\right)\cdot
tr\left(F_{np}^2(y) + ...\right)>_{\epsilon}\Bigr\},
\label{gen}
\eeqn
up to the {\it quadratic} order of the dilaton. This expression contributes
to the renormalization of the unity operator which couples to the volume
element on the D-brane.

   In eq. \eq{gen} the correlator $<tr\left(F_{lm}^2 + ...\right)\cdot
tr\left(F_{np}^2 + ...\right)>_{\epsilon}$ is a regularized version of
the $<tr\left(F_{lm}^2 + ...\right)\cdot tr\left(F_{np}^2 +
...\right)>$.  From \eq{boundary},\eq{regac} and \eq{gen} we get that
\cite{GuPoKl98,Wit98}:

\beqn
<tr\left(F_{lm}^2(x) + ...\right)\cdot tr\left(F_{np}^2(y) +
...\right)>_{\epsilon} \sim \frac{N^2}{\left(\epsilon^2 +
|{\vec x} - {\vec y}|^2\right)^4} - \nonumber \\ -
2 \epsilon^2 \frac{N^2}{\left(\epsilon^2 + |{\vec x} - {\vec y}|^2\right)^5},
\label{regsc}
\eeqn
which is natural if considered as the "point splitting" in the extra (fifth)
dimension. This regularization scheme can be formulated via an inclusion of
non-local terms into the action \eq{genSYM}, which can be expanded in powers
of $\epsilon^2$.

   Now we will use those considerations to compute counter-terms which
renormalize the dilaton. Although within the theory from eq. \eq{genSYM} the
dilaton does not get renormalized\footnote{Because fermionic loops perfectly
cancel that of bosons in the $\cN = 4$ SYM theory.  This is true even if we
consider the dilaton as some background {\it non-constant} field.}, in the
LHS of \eq{boundary} there are higher derivative terms included in $\cO^j$
operators. They can lead to the deformation of the dilaton.

    It happens, though, that the next to leading term from the non-Abelian
variant of the \eq{BI} -- $tr\left\{F^4 - \frac14 \left(F^2\right)^2\right\}$
-- gives {\it no} contribution to the
dilaton renormalization\footnote{It gives, however, non-zero contribution to
the renormalization of the unity operator \cite{HaGuKl}.}. It is not a
coincidence. All terms in \eq{BI} coming from the disc topology also should
{\it not} give such a contribution. In fact, they are of the order of $N^2$,
if $N\to\infty$.  Then, as one can estimate, if those terms could renormalize
the dilaton, we would get contributions of the order of $N^4$.  Which
contradicts to all our expectations from the large $N$ limit \cite{t'Ho74}.
Hence, we can accept that there is no
deformation of the dilaton coming from the disc topology and, in particular,
from the leading action presented in \eq{BI}.

   There are, however, terms coming from higher topologies.
The obvious leading term (in powers of derivatives) among them is given by:

\beq
\cS_2 = \frac{a \epsilon^4}{N^2}\left[tr \left(F_{mn}^2 + ...\right)\right]^2.
\label{termin}
\eeq
Here $a$ is some constant irrelevant for our discussion below.
The expression \eq{termin} is of the order $N^0$, as $N\to\infty$. Thus, it
can give a proper counter-term ($\sim N^2$) which renormalizes the dilaton.

   So, we add $\cS_2$ to the action in eq. \eq{genSYM}. After that,
we represent the super-gauge fields as $A_m = \bar{A}_m + \cA_m$ and etc.
for the super-partners. Where $\cA_m$ are quantum fluctuations over the
background $\bar{A}_m$. Then, we expand the action from \eq{genSYM},
\eq{termin} in powers of the $\cA_m$ and integrate over them at one loop. As
we have already mentioned, from the action in \eq{genSYM} there do not appear
terms renormalizing the dilaton. For there is a perfect cancellation of
the fermionic and bosonic loops. However, because of the presence of the
$\cS_2$ term, we get several counter-terms. Among them we are interested in
the {\it only one} which contributes to the dilaton renormalization at the
{\it linear} dilaton order.  It is given by the second expression in the
equation:

\beqn
S_{eff}(\varphi_0, \bar{A}) = \frac{1}{g_{ym}^2}
\int d^4 x~~\varphi_0 \cdot tr
\left(\bar{F}_{mn}^2 + ...\right) + \nonumber \\ +
\frac{a \epsilon^4}{g_{ym}^2 N^2} \int d^4 x \int d^4 y~~\varphi_0 (y)
\cdot tr \left(\bar{F}_{mn}^2(x) + ...\right)
<tr\left(\cF_{lp}^2(x) + ...\right)\cdot tr\left(\cF_{rt}^2(y) + ...
\right)>_{\epsilon}.
\eeqn
Here $\bar{F}_{mn}$ and $\cF_{mn}$ are the gauge field strengths of the
vector potentials $\bar{A}_m$ and $\cA_m$, respectively.

   To find the counter-term in question, we take the expression
\eq{regsc} for the correlator $<...>_{\epsilon}$.  Now if the dilaton
$\varphi_0$ is an arbitrary constant this counter-term is a trivial constant.
While in the case when the dilaton is excited and equals to $\varphi_0 = e^{i
{\vec k}{\vec x}}$ we get a renormalization. After the redefinition ${\vec y}
- {\vec x} \to {\vec y}$ of the integral over ${\vec y}$ we obtain:

\beqn
S_{eff}(\varphi_0, \bar{A}) = \frac{1}{g_{ym}^2}
\int d^4 x~~\varphi_0(x) \cdot tr
\left(\bar{F}_{mn}^2(x) + ...\right) +
\frac{a}{g_{ym}^2 N^2} \int d^4 x~~\varphi_0 (x)
\cdot tr \left(\bar{F}_{mn}^2(x) + ...\right) \times \nonumber \\ \times
\left[\epsilon^4 \int d^4 y~~\frac{e^{i{\vec k}{\vec y}}}{\left(|{\vec x} -
{\vec y}|^2 + \epsilon^2 \right)^4} - 2 \epsilon^6
\int d^4 y~~\frac{e^{i{\vec k}{\vec y}}}{\left(|{\vec x} -
{\vec y}|^2 + \epsilon^2 \right)^5}\right]. \label{integrals}
\eeqn
It is easy to calculate the integrals in the second line of this
formula because they are related to the $\cK_2(k\epsilon)$ Bessel function
and its derivatives.

   The requirement of the renormalization group invariance:

\beq
\epsilon\diff_{\epsilon}\left\{\varphi_0(\epsilon,x)\left[ 1 + \Phi (\epsilon)
\right]\right\} = 0,
\eeq
where $\Phi (\epsilon)$ schematically represents the counter-term in question,
makes the dilaton to be dependent upon the scale $\epsilon$. Now,
calculating the integrals \eq{integrals}, expanding them to the fourth order
in the $k\epsilon$ and tuning the constant $a$, we get:

\beq
\varphi_0 (\epsilon,x) = e^{i{\vec k}{\vec x}} \left(1 + a_1 (k\epsilon)^2
+ a_2 (k\epsilon)^4 - \frac{1}{16}(k\epsilon)^4 \log{\frac{\gamma
k \epsilon}{2}} + ...\right). \label{expan}
\eeq
Here $\gamma$ is the Euler constant. Other terms in the expansion \eq{expan}
receive contributions from the higher derivative
corrections\footnote{For example, there can be terms like $\cS_n \sim
\frac{\epsilon^{4n-4}}{N^n}\left[tr\left(F^2_{mn} + ...\right)\right]^n,
\quad n\ge 3$. Which are coming from the next to leading topology and behave
as $N^0$, when $N\to\infty$.} to the action from eq. \eq{genSYM}. This
expression reproduces, up to the constants $a_1$ and $a_2$, the expansion of
\eq{ren} over $(kz)^2$, when $\epsilon = z$.  Unfortunately, at present we
are not able to recover from the renormalization group the exact values of
the coefficients $a_1$ and $a_2$. In fact, they stand in front of the contact
terms which get contributions from {\it all} higher corrections.  Moreover,
they can be altered via introduction of local boundary terms, which are yet
undetermined, into the action on the RHS of \eq{boundary}. The boundary terms
would change the value \eq{regsc} of the correlator $<tr\left(F_{mn}^2(x) +
...\right)\cdot tr\left(F_{pq}^2(y) + ...\right)>_{\epsilon}$.  But it is
important that the {\it universal} term with logarithm does not receive any
other contributions. Also, because of the SUSY we expect only {\it linear}
logarithm corrections which is in agreement with the expansion of \eq{ren}
over the $(kz)^2$.

4. We may conclude that after account of the higher derivative terms from
the LHS of the eq. \eq{boundary}, the dilaton becomes dependent on the
normalization scale. Very probably this dependence is governed by the
classical SUGRA equations of motion.

  At this point one may ask at least the following question:
why the renormalization group equations are of the second order rather than of
the first one?  The answer possibly is as follows. We expect that
the renormalization group equation for the dilaton depends {\it only} upon
the unity operator and vise versa. If this is true, looking for the equation
only in terms of the dilaton, one would find it to be of the second order.
This possibility is supported by the fact that the second solution of the
modified Bessel equation \eq{equ} has a good property to correspond to the
unity operator. In the limit $kz \to 0$ the solution behaves as $\sim z^4$,
which compensates the scale dependence of the D-brane space-time volume. As
it should be in the conformal field theory \cite{Anton}.

5. Author is indebted for valuable discussions to A.~Losev, A.~Mironov,
A.~Gorski, A.~Morozov, I.~Polyubin and especially to A.~Gerasimov.
Author also wants to thank A.~DiGiacomo, M.~Mintchev, B.~Lucini, L.~Montesi
and P.~Barletta for hospitality at the Pisa University where this work was
completed.  This work was partially supported by RFFI 97-02-17927 and INTAS
96-538.



\begin{thebibliography}{50}
\bibitem{Anton} Author is indebted to A.~Gerasimov for discussions
on this point.
\end{thebibliography}
\end{document}